\documentclass[9pt]{article}
\usepackage[fleqn]{amsmath}
\usepackage{spconf,graphicx}
\usepackage{enumitem}
\setlist{nosep, leftmargin=14pt}

\usepackage{amsfonts}
\usepackage{array}
\usepackage{textcomp}
\usepackage{stfloats}
\usepackage{url}
\usepackage{verbatim}
\usepackage{tikz}
\usepackage[english]{babel}
\usepackage[utf8]{inputenc}
\usepackage{algorithm}
\usepackage{algpseudocode}
\usepackage{booktabs}
\usepackage{xparse}
\usepackage{xcolor}
\usepackage[normalem]{ulem}
\usepackage{bbm}
\usepackage{float}
\usepackage{soul} 
\usepackage{sidecap, caption}
\usepackage[font={small}]{caption} 
\usepackage[nodisplayskipstretch]{setspace}

\DeclareMathOperator*{\argmax}{argmax} 
\numberwithin{equation}{section}

\newcommand{\E}{\mathbb{E}}
\newcommand{\trace}{\text{trace}}
\newcommand{\p}{\text{p}}

\usepackage{titlesec}
\titleformat{\section}{\bfseries\centering\MakeUppercase}{\thesection}{1em}{}
\titlespacing*         
    {\section}         
    {0pt}              
    {0.25\baselineskip} 
    {0.25\baselineskip} 

\titlespacing*        
    {\subsection}     
    {0pt}             
    {0.25\baselineskip}
    {0.25\baselineskip}

\algdef{SE}[SUBALG]{Indent}{EndIndent}{}{\algorithmicend\ }%
\algtext*{Indent}
\algtext*{EndIndent}
\newcommand{\rvc}[2]{
\protect\textcolor{red}{#1}%
    \IfNoValueF{#2}{\protect\footnote{#2}%
    }%
}

\title{Unsupervised particle sorting for cryo-EM using probabilistic PCA}
\name{Gili Weiss-Dicker $^{\star}$ \qquad 
      Amitay Eldar $^{\dagger}$ \qquad 
      Yoel Shkolinsky$^{\dagger}$ \qquad 
      Tamir Bendory$^{\star}$}

\address{$^{\star}$ School of Electrical Engineering,
	Tel Aviv University 
    \\
    $^{\dagger}$ Department of Applied Mathematics, School of Mathematical Sciences, Tel Aviv University
    } 
\begin{document}
%
\maketitle
\begin{abstract}
Single-particle cryo-electron microscopy (cryo-EM) is a leading technology to 
resolve
the structure of molecules. 
Early in the 
process, the user detects 
potential
particle images in the raw data.  
Typically, 
there are
many false detections as a result of high levels of noise and contamination. 
Currently, removing the false detections requires human intervention to sort the hundred thousands of images. 
We propose a 
statistically-established unsupervised algorithm to remove non-particle images.
We model the  particle images as
a
union of low-dimensional subspaces,
assuming non-particle images are arbitrarily scattered in the high-dimensional space. 
The algorithm is based on an extension of the probabilistic PCA framework to robustly learn a non-linear model of union of subspaces. 
This 
provides a flexible model for cryo-EM data,
and allows to automatically remove images that correspond to pure noise
and contamination. Numerical experiments 
corroborate  the effectiveness of the sorting algorithm.
\end{abstract}

\section{Introduction}

\begin{figure*}[b]
\vspace{-4 mm} 
    \centering
    \centerline{\includegraphics[width=0.8\linewidth]{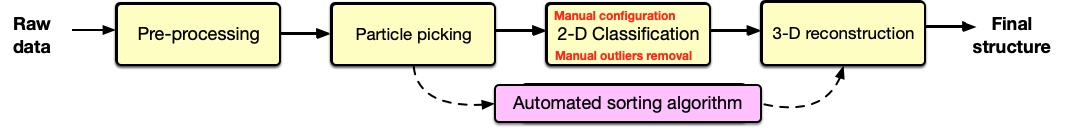}}
    \caption{A typical data processing pipeline for single-particle cryo-EM. The main contribution of this work is the fully automated sorting block, aiming to replace the need of an expert practitioner involved in the sorting step.}
    \label{fig:pipeline}
\end{figure*}

Single-particle cryo-electron microscopy (cryo-EM) is an emerging technology to determine the 
structure of molecules.
In the cryo-EM process, the acquired  ``raw data'' image, called a micrograph,  
 contains a few dozens of 2-D tomographic particle projection images with unknown random orientations and locations. 
 The micrograph suffers from low signal-to-noise ratio (SNR), as low as $\frac{1}{100}$.
 Typically, it also contains undesired contamination.
For the purpose of this paper, the pixels in a micrograph can be broadly divided into three categories: regions of particles with additive noise, regions of contamination, and regions of noise only.

\begin{figure}
  \begin{minipage}[c]{0.63\columnwidth}
    \includegraphics[width=\columnwidth]{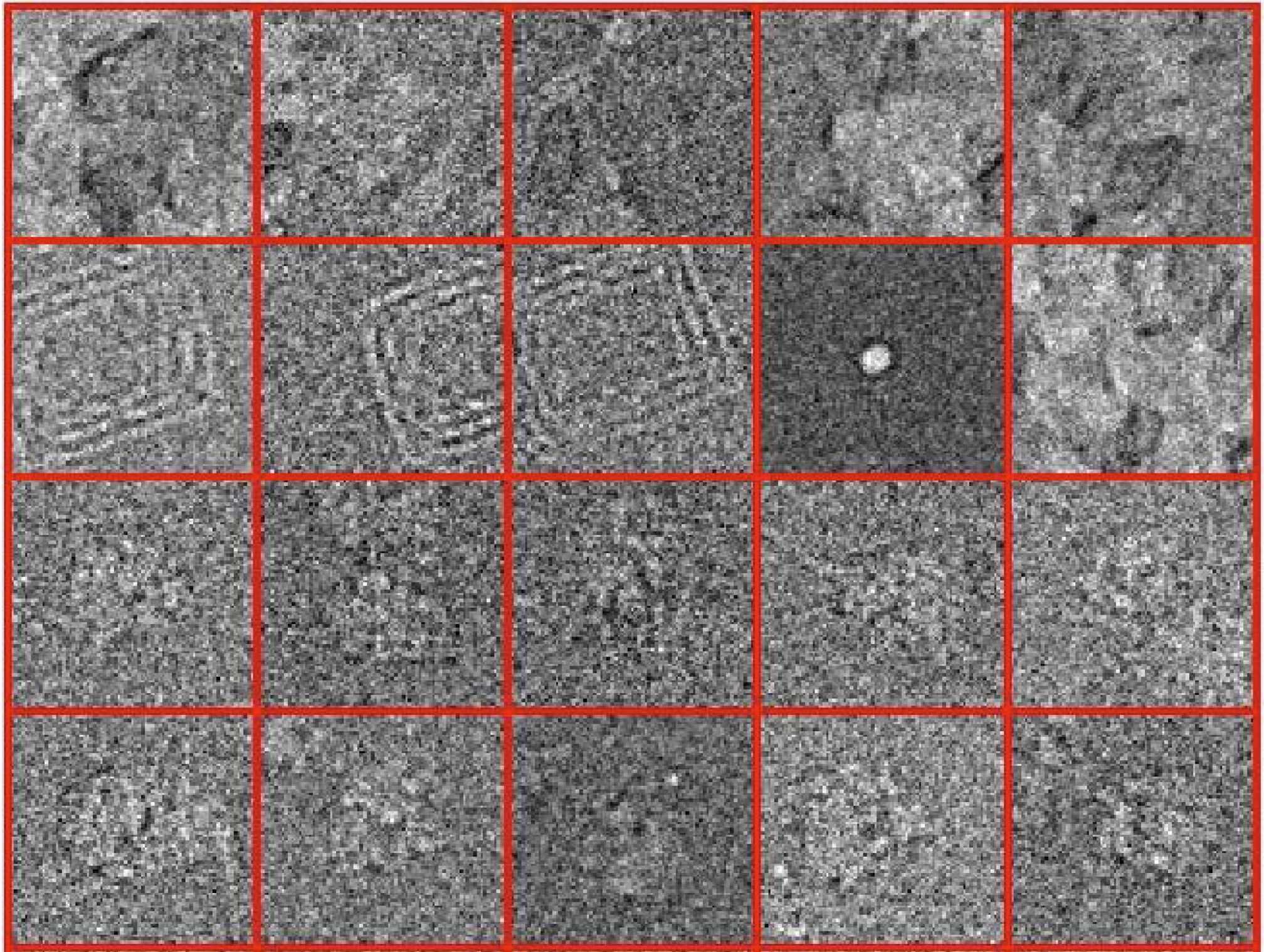}
  \end{minipage}\hfill
  \begin{minipage}[c]{0.34\columnwidth}
    \caption{Highest scoring 
    images
    by the KLT picker~\cite{eldar2020klt} on Plasmodium falciparum 80S ribosome~\cite{wong2014cryo}. 
    Two bottom rows consist of particle images, while the others are contamination images.
    }
    \label{fig:particles}
  \end{minipage}
  \vspace{-6 mm} 
\end{figure}

During the 
cryo-EM workflow, 
particle images are 
detected and extracted
from micrographs in a process called particle picking~\cite{bendory2020single,singer2020computational}. 
The
extracted images are the individual particles within each micrograph. 
If only particles were picked, the
images chosen by the particle picker would have been used to construct the 3-D molecular structure.
Figure~\ref{fig:pipeline} illustrates a schematic sequence of computational steps 
typically used to convert the raw data into 3-D molecular structures. 
While many particle picking algorithms were developed, e.g., ~\cite{scheres2015semi,wang2016deeppicker,
eldar2020klt}, 
due to very low  SNR levels they  result in
contamination 
and pure noise images 
picked along with the particle images. 
Typical images chosen by a picking algorithm can be seen in Figure~\ref{fig:particles}.

A common approach to remove non-particle images, called ``2-D classification,'' is semi-automatic and involves an expert practitioner; 
it relies heavily on subjective criteria that are neither consistent nor reproducible among different users.  
We propose an automatic 
and statistically-established unsupervised algorithm to remove non-particle images from the data. 
Specifically, we assume that all particle images approximately lie on a union of subspaces, whereas the non-particle images are scattered  in the high-dimensional space. 
Similar parsimonious  models are ubiquitous  in many signal processing tasks, and specifically in different stages of the 
cryo-EM computational pipeline~\cite{zhao2016fast,moscovich2020cryo}.  
The main computational tool in this work is an extension of principal component analysis (PCA).
PCA has been applied to cryo-EM data for several tasks~\cite{zhao2016fast, landa2017steerable}. However, 
PCA is limited since it learns a single subspace. 
We build on a maximum likelihood  formulation, called probabilistic PCA (PPCA)~\cite{tipping1999probabilistic, roweis1998algorithms, ghahramani1996algorithm}. 
In particular, we iteratively estimate the 
union of subspaces using an expectation-maximization (EM) algorithm, 
while sorting out  images that do not lie on the subspaces.

PPCA offers several attractive advantages over PCA. 
First, PPCA can be readily extended to multiple subspaces, leading to a nonlinear flexible mixture model.
Second, we work in the  dimension of the problem (i.e., the number of parameters that define the sought subspaces) in contrast to standard PCA that requires estimating the full covariance matrix. 


\section{Existing particle sorting solutions} \label{sec:previous_work}
\begin{figure}[htb]
\vspace{-2 mm} 
    \centering
    \includegraphics[width=5.7cm]{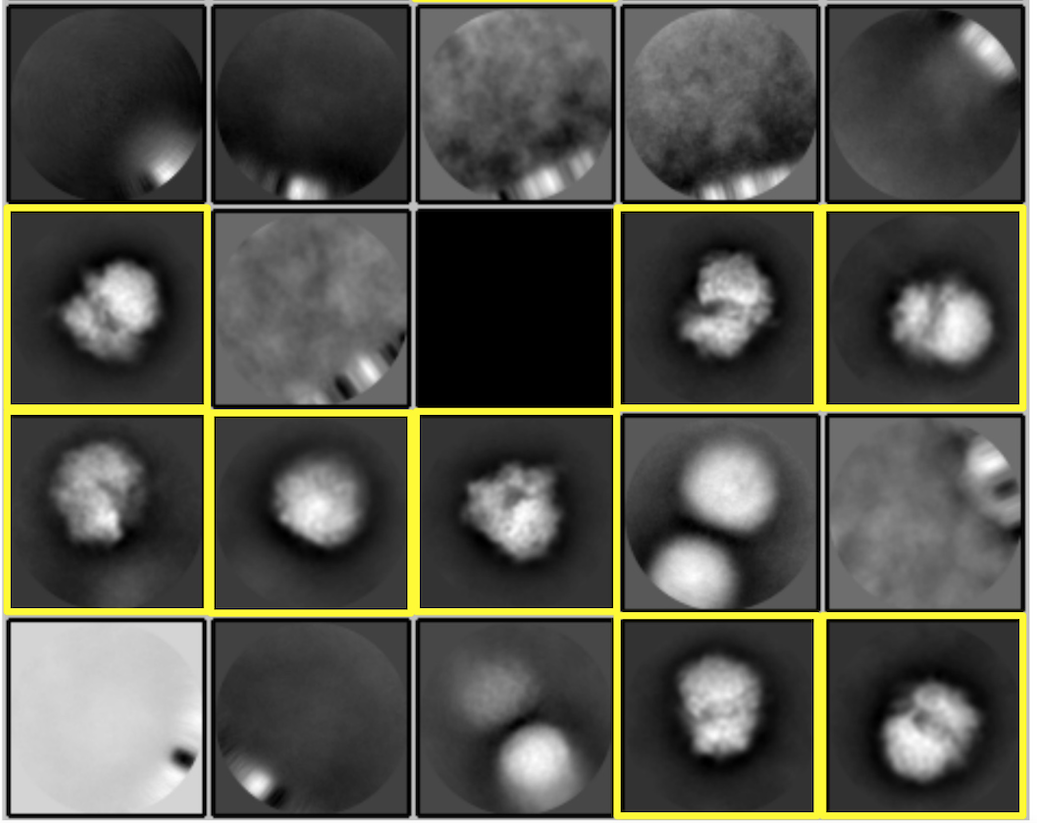}
    
    \caption{
    	    The current practice to remove non-particle images begins with  clustering similar images, a procedure called 2-D classification.
    	    The figure shows the results of RELION's 2-D classification tool~\cite{scheres2015semi}  on the Plasmodium falciparum 80S ribosome dataset~\cite{wong2014cryo}, picked using the KLT picker~\cite{eldar2020klt}.
            An expert manually chooses the particle classes (marked in yellow).
    	Then, the user re-runs the 2-D classification algorithm  on the images associated with the  selected classes.
    	This step is usually done multiple times to reach satisfying results.\\
    }

    \label{fig:2D}
\vspace{-2 mm} 
\end{figure}

Several algorithms have been proposed 
to automate the particle sorting step. 
In general, such algorithms can be divided into two main groups: semi-automatic and automatic. 
Since a typical cryo-EM dataset after particle picking contains hundreds 
thousands of images, it is unreasonable
to visually inspect each one of them. 
Thus, in a typical semi-automated sorting,
the images are clustered based on their  similarity using a 2-D classification tool~\cite{scheres2005maximum}.
An expert then visually inspects representative images of 
the
clusters and discards uninformative ones, as demonstrated in Figure~\ref{fig:2D}.

Almost all cryo-EM users remove the non-particle images using 2-D classification,
but a few automated methods have been proposed. 
One approach uses a statistical model of mixture of two Gaussians for the distribution of scores assigned during refinement~\cite{zhou2020unsupervised}; 
this method 
relies on the availability of a satisfying initial 3-D model of the investigated molecule..
Unfortunately, 
often, a reliable initial model is unavailable. 
Another approach uses 
transfer learning techniques 
to sort cryo-EM images
~\cite{li2021pickeroptimizer}. 
This method requires a labeled dataset, thus it reduces the amount of experts' manual work but it still  depends the user's subjective viewing point.
Another deep learning approach takes multiple picking algorithms' output and consolidates their results; 
this 
is time-consuming and demanding on computational resources~\cite{sanchez2018deep}. 
None of the existing works are unsupervised and self-sufficient.
In contrast, 
our proposed method
does not require any human intervention nor professional expertise involved.

\section{Unsupervised particle sorting}
\label{subsec:algo_summary}
Our approach is based on a nonlinear mixture model composed of factors of PPCA analyzers.
Our proposed algorithm consists of two main ingredients. The first  is the
union of subspaces model and the EM algorithm that aims to estimate the subspaces. The second is the sorting method, 
which removes images that are distant from the learned union of subspaces. 

\begin{figure*}[t]
\vspace{-5 mm} 
    \centering
    \includegraphics[width=0.82\textwidth]{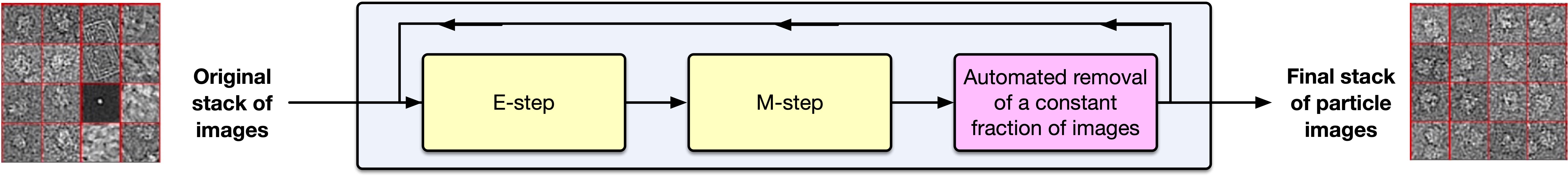}
    \caption{Schematic view of the sorting procedure. The EM iterations alternate between performing an E-step and an M-step; every several iterations we remove a constant fraction of the stack of images.}
    \label{fig:sort_scheme}
\end{figure*}

We fuse these two tools into one sorting algorithm: given a set of images picked by some particle picking algorithm, our goal is to find a subset of the particle images, while removing the contamination images.
The proposed EM method estimates the multiple subspaces simultaneously. 
As the EM algorithm is iterative, we concurrently sort out images which we consider ``far'' from the learned subspaces during the estimation process.
Namely, we discard the images which are most likely not to originate from the subspaces. 
Discarding images during each iteration
should improve the next learning iteration of subspaces, as the outliers---which do not lie on the subspaces---are discarded.
Thus, we gain two birds with one stone. 
First, the estimation process is improved, assuming we removed outlier images (i.e., non-particle images). Second, we reduce the  computational burden as the data set contains fewer images.
A scheme of the algorithm is presented in Figure~\ref{fig:sort_scheme}. 

We begin by  mathematically formulating the  image generative model.
Next, we explain how we estimate the sought subspaces 
via the
EM algorithm. Ultimately, we present our approach for online sorting. 
Numerical results are deferred to the next section.


\label{sec:model}


\subsection{Image formation model} 
\label{subsec:image_model} 
We assume that the particle images lie in a low-dimensional (non-linear) union of $M$ unknown subspaces, whereas the non-particle images are arbitrarily scattered in the high $L$-dimensional space. 
Each image is assumed to originate from one of the subspaces. For subspace $1 \le m  \le M $, we define the mixture indicator variable $w_m \in {[0,1]}$, where $w_m = 1$ when the image was generated by subspace $m$.

We assume to acquire $N$ image realizations drawn from the model
\begin{small}
$
	y =  C_{w_m} x +\mu_{w_m} + \varepsilon,
$
\end{small}%
where $C_1,\ldots,C_M\in \mathbb{R}^{L \times K}$ are matrices that span the sought subspaces,  $\mu_1,\ldots,\mu_M \in \mathbb{R}^{L \times 1}$ are mean vectors that are the centers of the subspaces, 
$K << L$,  and $x \in \mathbb{R}^{K \times 1} $ are the low-dimensional coefficients, lying in the union of subspaces, which are normally distributed with zero mean and unit covariance.
The additive noise is normally distributed
$\varepsilon \sim \mathcal{N}(0, \sigma^2 I_L )$.
The vector $\pi=(\pi_1,\ldots,\pi_M)^T$ denotes a distribution of the particles over the $M$ subspaces so that  $\pi_m = \p(w_m=1)$, where $\p(\cdot)$ denotes  probability densities. Furthermore, 
 $\varepsilon$, $w_m$ and $x$ are  independent.

 We are interested in estimating the 
 model
 parameters and~$\pi$, 
 while $x_1,\ldots,x_N$ are treated as nuisance variables. We also estimate the noise variance~$\sigma^2$. 
 Accordingly, the generative model reads 
 $
 	\p(y) =\sum_{m=1}^M \int \p(y|x,w_m)\p(x)\p(w_m)dx,
  $
with $x \sim \mathcal{N}(0,I_K)$.

We use the prolate spheroidal wave functions (PSWFs)~\cite{slepian1964prolate,landa2017steerable} basis to reduce dimensionality of the input images, but other bases may be used instead.
In the sequel, with a slight abuse of notation, we refer to the PSWF coefficients of the images 
as the images.

\subsection{EM algorithm} 
\label{subsec:EM} 


Expectation-maximization (EM) is an iterative method to
maximize the likelihood function of statistical models
that include nuisance variables~\cite{dempster1977maximum}. 
In our problem, given the observations $y_1,\ldots,y_N$, we wish to estimate the subspaces' parameters, the distribution over the subspaces,  and the noise variance, that maximize  the  likelihood function.
We denote all parameters by $\theta$.
The nuisance variables of the model---namely, unknowns whose estimation is not the aim of the task---are the coefficients $x_1, \ldots, x_N$ and the indicators $w_m$ associated with each measurement. 

Let
\begin{small}
\begin{equation} \label{eq:theta_t}
    \vspace{-3 mm} 
	\theta_t = \left\{ C_1^{(t)}, ..., C_M^{(t)}, \mu_1^{(t)}, \ldots,\mu_M^{(t)}, \pi^{(t)}, {\sigma^2}^{(t)} \right\}
\end{equation}
\end{small}%
be the estimate of the model parameters 
at the  $t$-th EM iteration. 
At each EM iteration, we wish to maximize the expected complete log-likelihood, given by 
\begin{small}
		$
		Q(\theta|\theta_t) = \E_{x,w|y,\theta_t} \left [\mathcal{L}(\theta;y,x,w) \right ],
		$
\end{small}%
where 	$\mathcal{L}( \theta; y,x,w)$ is the 
complete-data log-likelihood of the generative model. Explicitly, it is given, up to a constant,  by
\begin{small}
\begin{flalign}
 \vspace{-12 mm} 
    \label{eq:llh}
        Q(\theta|\theta_t) \nonumber
	& = \sum_{i=1}^{N} \sum_{m=1}^{M} \E_{x,w|y,\theta_t} [w_m \log{ \p(y_i| x_i,w_m) \p(x_i)}]     \nonumber \\
        & =  -\frac{L}{2} \log{ {\sigma^{2}}^{(t)} } - \frac{1}{2 {\sigma^{2}}^{(t)}}  \sum_{i,m}  h_{i,m}  \bigg(\| y_i - \mu_m^{(t)}  \|^2\nonumber\\
        & + \trace( {C_m^{(t)}}^{T} C_m^{(t)}  \E[x_i x_i^T |y_i,w_m] ) \\
        & + 2 (\mu_m^{(t)}  - y_i)^TC_m^{(t)}  \E[x_i |y_i,w_m]\bigg), \nonumber
        \vspace{-10 mm} 
\end{flalign}
\end{small}
where 
\begin{equation}
	\vspace{-2 mm} 
	\label{eq:h_im}
		h_{i,m} = \E(w_m| y_i). 
\end{equation}  

Each EM iteration consists of two steps.
For the first step, the E-step, we 
calculate the expectations of the nuisance variables that appear in the Q function~\eqref{eq:llh}. 
It can be shown that 
\begin{small}
	\begin{flalign}
		\label{eq:ex}
		&\E[x_i| y_i,w_m] = B_m (y_i - \mu_m),  \\
		\label{eq:exx}
		&\E[ x_i {x_i}^T|y_i,w_m ] = I_K - B_m C_m^T 
		+ B_m (y_i - \mu_m) (y_i - \mu_m)^T B_m^T, 
		\vspace{-8 mm} 
	\end{flalign}
\end{small}%
where
\begin{small}
$
B_m = C_m^T(C_m C_m^T + \sigma^2 I_L)^{-1} \in \mathbb{R} ^{K \times L}.
$
\end{small}%

For the second step of the EM algorithm, the M-step, we 
maximize the expected log-likelihood, namely, solve
\begin{small}
$
    \theta^{(t+1)} = \argmax_{\theta} Q(\theta|\theta_t),
$
\end{small}%
assuming the expectations of the nuisance variables are known. 
The maximum is attained as the solution of the linear system of equations:
\begin{small}
\begin{flalign}
 \vspace{-5 mm} 
    \label{eq:c_m}
  {C_{m}^{(t+1)}}  = 
    & \left( \sum_{i} h_{i,m}(y_i - {\mu_{m}^{(t+1)}}) \E[ x_i^T |y_i,w_m ] \right) \\
    & \left(\sum_{i} h_{i,m}  \E [ x_i x_i^T|y_i,w_m  ]  \right) ^{-1}, \nonumber
     \vspace{-8 mm} 
\end{flalign}
\begin{equation}
    \label{eq:mu_m}
    {\mu_{m}^{(t+1)}}  = \frac{1}{N} \sum_{i}h_{i,m}  (y_i - {C_{m}^{(t+1)}} \E[x_i | y_i,w_m ])
    ,
     \vspace{-5 mm} 
\end{equation}
\begin{flalign}
    \label{eq:sigma_m}
	{\sigma^2}^{(t+1)}
	= & \frac{1}{L} \sum_{i,m}  h_{i,m}^{(t)}  ( 
	\| y_i - \mu_m^{(t+1)} \|^2 \nonumber\\
	& + \trace( C_m^{T(t+1)} C_m^{(t+1)} \E[x_i x_i^T |y_i,w_m]) \\
	& + 2 (\mu_m^{T(t+1)} - y_i^T )C_m^{(t+1)} \E[x_i 
	|y_i,w_m] ). \nonumber
  \vspace{-2 mm} 
\end{flalign}
\end{small}%

\begin{table}[htb]
    \centering
    \caption{Simulated data results of Section~\ref{sec:simulated}, $K_{\text{total}} = 60$.}
    \resizebox{\columnwidth}{!}{%
    \begin{tabular}{cccccc}
    \toprule
      & Raw Data & 1 Subspace & 2 Subspaces & 3 Subspaces & sPCA \\
    \midrule
    \% Particles &83.33 &87.89 &89.25 &\textbf{91.04} &90.1\\
    \midrule
    \% Outliers   &8.33 &3.38 &3.95 &\textbf{2.9}  &9.6\\
    \midrule
    \% Noise      &8.33 &8.73 &6.8 &6.06  &0.84\\
    \bottomrule
    \end{tabular}}
    \label{tab:simulated} 
\vspace{-9pt}
\end{table}

\subsection{Online sorting}
\label{subsec:sorting}

Assuming  the particle images lie in the learned subspaces, 
for
an image $y_i \in \mathbb{R}^{L \times 1}$, and a subspace $C_m \in \mathbb{R}^{L \times K} $, 
the sorting factor is defined by
    \begin{small}
    \begin{equation}
        \label{eq:sf}
        \text{SF}_m(y_i) = \frac{ \text{projection error}}{ \text{projected energy}}  = \frac{ \|  C_m C_m^T y_i - y_i \|_{\text F}^2 }{ \|  C_m C_m^T y_i \|_{\text F}^2 },
     \vspace{-3 mm} 
    \end{equation}
    \end{small}%
where $C_m C_m^T y_i$ is the projection of $y_i$ onto the subspace spanned by $C_m$. 
A large 
value indicates the image is far from the subspace.

In practice, after every constant number of $P$  iterations of the EM algorithm, we calculate the SF for each pair of an image and a subspace, and sort the images by summing up the factor value for all  subspaces (per image). Then, at every sorting step, we remove a constant fraction, denoted by $\alpha$, of the images in the stack.  The algorithm is outlined  in Algorithm~\ref{alg:algo}.

\section{Numerical Experiments} 
\label{sec:results}
In this section, we describe 
numerical results 
on synthetic and experimental data sets. 
 The 
 code 
 to reproduce all experiments 
is 
 publicly 
available at \url{https://github.com/giliw/particle_sorting}. 

\subsection{Simulated cryo-EM data}
\label{sec:simulated}
\vspace{-1 mm} 
 We generated  
 50K
 tomographic projections of $71 \times 71$ pixels of 
the 
80S ribosome, 
available at the Electron Microscopy Data Bank
(\url{www.ebi.ac.uk/emdb/EMD-2275}), using the ASPIRE 
package (\url{www.spr.math.princeton.edu}). We added additive white Gaussian noise, resulting in SNR = $\frac{1}{10}$.
We added random small shifts to the images. 
We also added outlier  images---representing  contamination in the micrographs---modeled as random cuts of Matlab's ``camera-man'' image, which were $8.33\%$ of the images. 
In addition, we added pure noise images which also comprised $8.33\%$ of the images. 
Throughout all the experiments, 
every $P = 6$ EM iterations
we  sort out $ \alpha = 5\%$ of the images. 

We run the unsupervised sorting algorithm with 1, 2 and 3 subspaces, where the total dimension was fixed to $K_{\text{total}}=60$. 
For example, when we estimated two subspaces,  the dimension of each subspace was set to 30. 
We tested our sorting algorithm against a sorting algorithm based on 
steerable 
PCA
(sPCA)~\cite{zhao2016fast}. This algorithm
takes $K_{\text{total}}$ eigenimages of  sPCA (corresponding to the largest eigenvalues), and then projects the images to that linear subspace. Finally, it sorts the reconstructed images based on the relative energy of the images with respect to the original images, and takes those with the highest value.
We compare against this algorithm since it is a linear algorithm, and is based on a popular cryo-EM practice.
  
Table~\ref{tab:simulated} presents the 
results. 
The original simulated images are referred to as  ``Raw Data''. 
We see that three subspaces were best
at sorting out the outliers, improving the $83.333 \%$ in the original dataset to $91.04 \%$ particle images.
Our numerical experiments indicate that more than three subspaces do not improve the results.
Although sPCA seems to achieve a similar particle images percentage, it almost did not discard  outlier images. 
In particular, we note that sPCA works well in removing the pure noise images, but this is a much easier problem that can be solved using standard statistical tests for detecting Gaussian noise, and is not the focus of this work.


\begin{table*}
\vspace{-11 mm} 
\begin{minipage}{\columnwidth}
 
    \centering
    \caption{Sorted tagged EMPIAR-10028 dataset, $K_{\text{total}} = 60$.}
    \resizebox{\columnwidth}{!}{%
    \begin{tabular}{cccccc}
        \toprule
         & Raw data & 1 Subspace & 2 Subspaces & 3 Subspaces & sPCA \\
         \midrule
        \% Particles   & 97 &	97.75	&	\textbf{98.2}	&	97.9 & 97.45
        \\
        \midrule
        \% of the outliers sorted    & - & 31.82 &  \textbf{45.45} &  36.36 &  22.73 \\
        \midrule
        $\#$ Images    & 742 & 667 & 667 & 667 & 667\\
        \bottomrule
    \end{tabular}}
    \label{tab:tagged_experimental} 
  \end{minipage}
  \begin{minipage}{\columnwidth}
     \centering
    \caption{Sorted EMPIAR-10028 dataset, $K_{\text{total}} = 60$.}
    \resizebox{\columnwidth}{!}{%
    \begin{tabular}{cccccc}
        \toprule
         & Raw data & 1 Subspace & 2 Subspaces & 3 Subspaces & sPCA \\
        \midrule
        \% Particles   & 68.2      & 86.4   &\textbf{95.4}   & 89.1  & 91.3
        \\
        \midrule
        \% Outliers    & 31.6      & 13.5   &\textbf{4.3} & 10.7 & 8.6\\
        \midrule
        $\#$ Images    &20.5K/30K & 17358/20K & 19187/20K & 17923/20K & 18355/20K\\
        \bottomrule
    \end{tabular}}
    \label{tab:experimental} 
\end{minipage}\hfill 
\vspace{-5 mm}
\end{table*}

\subsection{Experimental cryo-EM data}
\label{subsec:cryoEM} 
 \vspace{-1 mm} 
The performance of our algorithm is assessed on the EMPIAR-10028 \cite{wong2014cryo} data set,
picked using the KLT picker~\cite{eldar2020klt}. 
For the next two experiments, we used the 
competing algorithm as described in Section~\ref{sec:simulated}. We evaluated our sorting algorithm with  1, 2, and 3 subspaces. 

In order to draw reliable conclusions, the first experiment uses 
manually tagged images. 
 To this end, 
 we eyeballed the clearer particle and contamination images from dozens of micrographs.
 The tagged dataset consists of 720 
 particles
 and 22 contamination images, referred to as  ``Raw Data'' in Table~\ref{tab:tagged_experimental}.
Our algorithm left 667 images out of the 742 images. 
Same images were discarded by the sPCA competitor algorithm.
In Table~\ref{tab:tagged_experimental} we see that two subspaces resulted in the best sorting results, removing 45.45\% of the outlier images, whereas sPCA removes only 22.73\%.
Our algorithm outperforms the sPCA even when we use a single subspace.

The second experiment 
was conducted using 
30K
images.
Having no ground truth labels, to evaluate algorithm performance,
we executed the same procedure a practitioner would have executed without an automated sorting algorithm.
Thus, we applied the RELION 2-D classification algorithm~\cite{scheres2015semi} to the 
dataset.
This
algorithm maps the images into clusters, some of which are clearly particle clusters. 
As a benchmark, 
the results of the 2-D classification on the originally picked images---without any sorting---are referred to as   ``Raw Data'' in~Table \ref{tab:experimental}.
Table~\ref{tab:experimental} shows that the sorting obtained using two subspaces achieved the best performance at sorting
real contamination images from the dataset. Beginning with $ 68.2 \%$ of particle images  in the raw data, the automatic sorting algorithm  has improved the ratio of particle images to $95.4 \%$.

\section{Discussion}
\label{sec:conclusions}
 \vspace{-1 mm} 
We have presented an unsupervised particle sorting algorithm
for cryo-EM. 
This is the first attempt at designing a completely unsupervised algorithm for sorting non-particle images, which is a major, time-consuming problem in cryo-EM. Our algorithm can be applied after any picking algorithm and be integrated into any cryo-EM pipeline.
Our numerical experiments provide strong indications that this approach might be useful for cryo-EM users. 
To improve the results, we intend to integrate statistical priors into the EM algorithm, 
such as
biological priors~\cite{singer2021wilson}.
As a future work, we intend to design data-driven techniques to determine hyper-parameters, such as the number of subspaces to be learned 
and how to decide how many images to sort out at each iteration. 
\vspace{-5 mm} 
\begin{algorithm}[b]
\scriptsize
\caption{Particle sorting using EM}\label{alg:algo}
\begin{algorithmic}
    \State {\textbf{Input:} 
        Stack of images $y_1,\ldots,y_N$, 
        $Q_0 \gets - \infty$ , $t \gets 0$,
        Initialization: $\theta^{(0)}~\eqref{eq:theta_t}$,
        $P$ - number of  EM iterations between sorting steps,
        $\alpha$ - fraction of images to discard at each sorting step,
        $\epsilon$ - convergence threshold
}
    \State {\textbf{Output:}
        Stack of particle images
    }
    \While{$ \|Q_t -Q_{t-1}\| < \epsilon$} 
        \State \textbf{E-step:}  For all $i,m$ compute: 
           $h_{i,m}$~\eqref{eq:h_im},
           $\E[x_i | y_i, w_m]$~\eqref{eq:ex},
           \Indent 
           $\E[ x_i {x_i}^T | y_i,w_m]$~\eqref{eq:exx},
              and $Q$ (\ref{eq:llh}).
              \EndIndent
            
        \State \textbf{M-step:} For all $m$ compute:
              $\pi_m^{(t+1)}=\frac{1}{N}\sum_{i=1}^Nh_{i,m}$,
              \Indent 
              ${\sigma^2}^{(t+1)}$~\eqref{eq:sigma_m},
              solve a set of 
              linear equations 
              for 
              \EndIndent
              \Indent
              $C_m^{(t+1)}$~\eqref{eq:c_m} 
              and $\mu_m^{(t+1)}$~\eqref{eq:mu_m}.
              \EndIndent
        \State \textbf{Every $P$ EM iterations:} 
        \State Discard $\% \alpha $ of the images using the  
        SF method~\eqref{eq:sf}. 
        \State $ t = t+1$ 
    \EndWhile
\end{algorithmic}
\end{algorithm}
\clearpage

\section*{Compliance with ethical standards}
This is a numerical simulation study for which no ethical approval was required.
\section*{Acknowledgment}
The authors are grateful to Ido Hadi for insightful comments.
This research is support by the BSF grant no. 2020159, the NSF-BSF grant no. 2019752, the ISF grant no. 1924/21, 
the European Research Council (ERC) under the European Union's Horizon 2020 research and innovation programme (grant agreement 723991 - CRYOMATH), and by the NIH/NIGMS Award  R01GM136780-01. 





\bibliographystyle{IEEEbib}

\end{document}